\documentstyle[prl,aps,epsf]{revtex}
\begin{document}

%\draft
 \twocolumn[\hsize\textwidth\columnwidth\hsize\csname
 @twocolumnfalse\endcsname

\title {Excitation of phonons in a Bose-Einstein condensate by light 
scattering}

\author{D.M. Stamper-Kurn, A.P. Chikkatur, A. G\"{o}rlitz, S. Inouye, S. 
Gupta, D.E. Pritchard, and W. Ketterle}
\address{Department of Physics and Research Laboratory of
Electronics, \\
Massachusetts Institute of Technology, Cambridge, MA 02139}
\date{} \maketitle

\begin{abstract}
Stimulated small-angle light scattering was used to measure
 the structure factor of a  Bose-Einstein condensate in the phonon regime.
The excitation strength for phonons was found to be significantly 
reduced from that of free particles, revealing the presence of 
correlated pair excitations and quantum depletion
in the condensate.
The Bragg resonance line strength and line shift
agreed with predictions of a local density approximation.
\end{abstract}
]
\vskip1pc

Spectroscopic studies have been used to assemble a
complete understanding of the structure of atoms and simple 
molecules.
Similarly, neutron and light 
scattering have long been used to probe the microscopic
excitations of liquid helium \cite{grey78,soko95,nozi90,grif93},
and can be regarded as the spectroscopy of a many-body
quantum system.
With the experimental realization of
gaseous Bose-Einstein condensates, the spectroscopy of this new
quantum fluid has begun.

The character of excitations
in a weakly-interacting Bose-Einstein condensed gas depends on
the
relation between the wavevector of the excitation $q$ and the inverse
healing length $\xi^{-1} = \sqrt{2} m c_{s}/ \hbar$ which is the 
wavevector related to the speed of Bogoliubov sound $c_{s} = 
\sqrt{\mu/m}$ where $\mu = 4 \pi 
\hbar^{2} a n_{0} / m$ is the chemical potential, $a$
the scattering length, $n_{0}$ the condensate density, and $m$ the atomic 
mass.
For large wavevectors ($q \gg \xi^{-1}$), the excitations are particlelike with a quadratic
dispersion relation.
Excitations in the
free-particle regime have been accessed by near-resonant light scattering
\cite{sten99brag}.
For small wavevectors ($q \ll \xi^{-1}$), the gas responds
collectively and density perturbations propagate as phonons at the
speed of Bogoliubov sound.
Such quasi-particle excitations have been observed
at wavelengths comparable to the size of the trapped gas
\cite{earlyrefs} and thus were strongly influenced by boundary conditions.

In this Letter, we describe the use of Bragg spectroscopy to probe
excitations in the phonon regime.  Using two laser beams intersecting
at a small angle, excitations in a Bose-Einstein condensate were 
created with wavevector $q < \xi^{-1}$, thus optically 
``imprinting'' phonons into the gas.
The momentum imparted to the condensate in the form of quasi-particles 
was directly measured by a time-of-flight analysis.
This study is the first to explore phonons with wavelengths
much smaller than the size of the trapped sample, allowing a direct 
connection to the theory of the homogeneous Bose gas.
By direct comparison, we show the excitation of phonons to
be significantly weaker than that of free particles.  
This provides
dramatic evidence for the presence  of
correlated momentum excitations
in the many-body
condensate wavefunction.

In optical Bragg spectroscopy, an atomic sample is illumined by two
laser beams with wavevectors ${\bf
k}_1$ and ${\bf k}_2$ and a frequency difference $\omega$
which is much smaller than their overall detuning $\Delta$
from an atomic resonance.  The intersecting beams create a
periodic, traveling intensity modulation
$I_{mod}({\bf r}, t) = I \cos({\bf q} \cdot {\bf r} -
\omega t)$
where ${\bf q} = {\bf k}_1 - {\bf k}_2$.  Atoms exposed to this
intensity modulation experience a potential due
to the ac Stark effect of strength 
$V_{mod} = \hbar \Gamma^2 / 8 \Delta \times I_{mod} / I_{sat}$ \cite{cohe92}, from 
which they may scatter.
Here $\Gamma$ is the 
linewidth of
the atomic resonance and $I_{sat}$ the saturation intensity. 

The response of a many-body system to this perturbation can be 
evaluated
using Fermi's golden rule.
We express $V_{mod}$ in
second-quantized
notation
$\hat{V}_{mod} = V/2 \,  \left(\! \right. 
\hat{\rho}^{\dagger}({\bf q}) e^{-i \omega t}
+ \hat{\rho}^{\dagger}({\bf -q}) e^{+ i \omega t}
\left. \!  \right)
% \label{eq:vmod}
$
where 
$\hat{\rho}^\dagger ({\bf q}) = \sum_{k} \hat{a}^\dagger_{k+q} \hat{a}_{k}$ 
is the Fourier transform of the
atomic density operator at wavevector ${\bf q}$ and
$\hat{a}_k$ ($\hat{a}^\dagger_{k}$) is the destruction
(creation) operator for an atom with momentum $\hbar {\bf k}$.
For the ground state $|g \rangle$ with energy $E_g$,
the excitation rate per particle is then
\begin{eqnarray}
\frac{2 \pi}{N \hbar} \! \left( \!
\frac{V}{2} \! \right)^{\! 2} \! \sum_f | \langle f | \hat{\rho}^{\dagger}({\bf q})| g
\rangle|^2 \delta( \hbar \omega \! - \! (E_f \! - \! E_g)) & = &
2 \pi \omega_R^2 S({\bf q}, \omega) \nonumber
\end{eqnarray}
where excited states $|f\rangle$ have energy
$E_f$, $N$ is the number of atoms in the system,
and $\omega_R = V / 2 \hbar$ is the two-photon Rabi frequency.
Thus, light scattering directly measures the dynamical
structure factor, $S({\bf q}, \omega)$, which is the 
Fourier transform of density correlations in 
state $|g \rangle$ \cite{nozi90,java95scat}.  Integrating
over $\omega$ gives the static structure factor $S({\bf q}) =
\langle g | \hat{\rho}({\bf q}) \hat{\rho}^\dagger({\bf q}) | g
\rangle / N$.

In this work,
measurements were performed on both magnetically trapped and freely
expanding Bose-Einstein condensates of sodium.  
Condensates of  $\approx 10^7$ atoms 
were created by
laser and evaporative cooling and stored in a cigar-shaped magnetic trap with
trapping frequencies of $\omega_r = 2 \pi \times 150$ Hz
and $\omega_z = 2 \pi \times 18$ Hz in
the radial and axial directions, respectively \cite{mewe96bec}.  

\begin{figure}[ht]
 \epsfxsize=70mm
  \centerline{\epsfbox{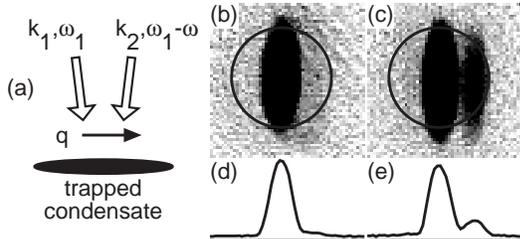}}
\caption{Observation of momentum transfer by 
Bragg scattering.
(a) Atoms were exposed to laser beams with wavevectors ${\bf k}_{1}$ and 
${\bf k}_{2}$ and frequency 
difference $\omega$, imparting momentum $\hbar {\bf q}$
along the axis of the trapped condensate.
Absorption images (b,c) after 70 ms time of flight show Bragg 
scattered atoms distinguished from the denser unscattered cloud
by their axial 
displacement.  Curves (d,e) show radially averaged (vertically in image) 
profiles of the 
optical density after subtraction of the thermal distribution.
The Bragg scattering velocity is smaller than the speed of 
sound in the condensate (position indicated by circle).
The response of trapped condensates (b,d) was much weaker than that
of condensates after a 
5 ms free expansion (c,e).
Images are 3.3 $\times$ 3.3 mm.}
   \label{images}
\end{figure}

The condensate was then exposed to two laser beams which
intersected at an angle of 14$^{\circ}$ and were aligned
symmetrically about the radial direction, so that the difference
wavevector ${\bf q}$ was directed
axially (Fig.\ \ref{images}a).
Both beams were derived from a common source, and then 
passed
through two acousto-optical modulators
operated with the desired frequency difference $\omega$,
giving both beams a detuning of 1.6 GHz to the red of the $|F=1\rangle
\rightarrow |F' = 0,1,2\rangle$ optical transitions.  Thus, at the optical
wavelength of 589 nm, the Bragg recoil velocity
was $\hbar q / m = 7.2$ mm/s, giving a predicted Bragg resonance 
frequency of $\omega_{q}^{0} = \hbar q^{2} / 2 m = 2 \pi \times 1.5$ 
kHz for free particles.  The beams were pulsed on at an
intensity of about 1 mW/cm$^2$ for a duration of 400 $\mu$s.
To suppress superradiant Rayleigh scattering \cite{inou99super}, both
beams were linearly polarized in the plane defined by the condensate 
axis and the wavevector of the light.

The effect of Bragg scattering on a trapped condensate was analyzed 
by switching off the magnetic trap 100 $\mu$s after the end of the light 
pulse, and allowing the cloud to freely evolve for 70 ms.
During the subsequent expansion, as the density of
the atomic cloud dropped, {\it quasi-particles}
in the condensate transformed into
{\it free particles} and were then imaged  by resonant absorption imaging 
perpendicular to the magnetic trap axis (Fig\ \ref{images}).
Bragg scattered atoms were distinguished from the 
unscattered atoms by their axial displacement.
The speed of Bogoliubov sound at the center of the
trapped condensate is related to the velocity of radial expansion 
$v_r$ as $c_{s} = v_{r} / \sqrt{2}$ \cite{cast96} 
($c_{s} = 11$ mm/s at $\mu / h = 6.7$ kHz as shown in Fig.\ 
\ref{images}).
Thus, comparing the axial displacement of the scattered atoms to 
the radial extent of the expanded condensate, one directly sees that 
the Bragg scattering recoil velocity
is smaller than the speed of sound in the trapped condensate,
i.e.\ the excitation in the trapped condensate occurs in
the phonon regime.

For comparison, Bragg scattering of free particles was studied by 
allowing the atomic sample to freely expand for 5 ms before 
application of a light pulse at equal intensity \cite{fallingfootnote}.
During the 5 ms expansion, the atomic 
density was reduced by a factor of
23 and the speed of sound by a factor of 5 from that of the trapped condensate.
Thus, Bragg scattering in the expanded sample occurred in the free-particle 
regime.

The momentum
transferred to the atomic sample was
determined by the average axial position in the time-of-flight 
images.
To extract small momentum transfers,
the images were first
fitted (in regions where the Bragg scattered atoms were absent)
to a bimodal distribution
which correctly describes the free expansion of
a condensate in the Thomas-Fermi regime, and of a thermal component 
\cite{kett99vartemp}.
The chemical potential $\mu$ of the trapped condensate was determined 
from the radial width of the condensate distribution \cite{cast96}.
The non-condensate distribution determined by the fit
was subtracted from the images before evaluating the momentum transfer.

\begin{figure}[ht]
\epsfxsize=50mm
 \centerline{\epsfbox{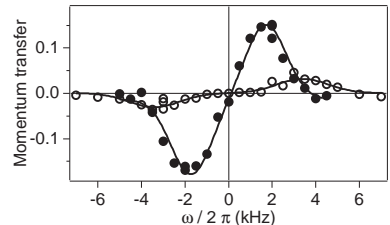}}
\caption{Bragg scattering of phonons and of free particles.
Momentum transfer per particle, in units of $\hbar q$,
is shown vs.\ the frequency difference $\omega / 2 
\pi$ between the two Bragg beams.
Open symbols represent the phonon excitation spectrum for a trapped 
condensate
at a chemical potential $\mu / h = 9.2$ kHz.
Closed symbols show the free-particle response of an expanded cloud.
Lines are fits to the difference of two Gaussian line shapes 
representing excitation in the forward and backward directions.}
   \label{spectra}
\end{figure}

By varying the frequency difference $\omega$, the Bragg
scattering spectrum was obtained for trapped and for
freely expanding condensates (Fig.\ \ref{spectra}).
The momentum transfer per atom, shown in units of 
the recoil momentum $\hbar q$, is anti-symmetric about $\omega = 0$ as
atoms are Bragg scattered in either the forward or backward direction,
depending on the sign of $\omega$ \cite{sqwfootnote}.

\begin{figure}[ht]
\epsfxsize=50mm
 \centerline{\epsfbox{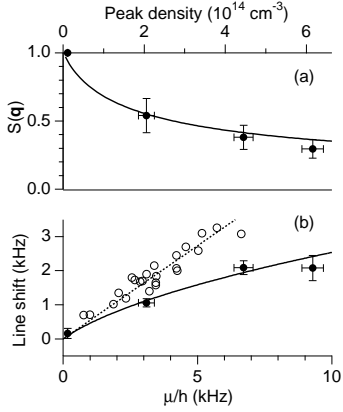}}
\caption{(a) Static structure factor $S({\bf q})$ and (b) shift 
of the line center from the free particle resonance.  $S({\bf q})$ is the ratio of 
the line strength at a given chemical potential $\mu$ to that observed for free particles.
As $\mu$ increases, the structure factor is reduced, and the Bragg resonance
is shifted upward in frequency.
Solid lines are predictions of a local-density approximation
using $\omega_{q}^{0} = 2 \pi \times$ 1.38 kHz.
Dotted line indicates a mean-field shift of $4 \mu / 7 h$ as 
measured in the free-particle regime (data from \protect\cite{sten99brag}
shown in open symbols).}
   \label{sqshift}
\end{figure}

From these spectra, we determined the total line strength and the
center frequency (Fig.\ \ref{sqshift}) by fitting the momentum transfer to
the difference of two Gaussian line shapes, representing excitation in
the forward and backward direction.
Since $S({\bf q}) = 1$ for free particles, we obtain the static 
structure factor as the  ratio of the line strengths for the trapped and the expanded 
atomic samples.
Spectra were taken for trapped condensates at three different
densities by compressing or decompressing the condensates in the magnetic
trap prior to the optical excitation.

The Bragg resonance for the expanded cloud was centered at 1.54(15) 
kHz, in agreement with the expected 1.5 kHz response for free 
particles \cite{residualfootnote}, with an rms-width of 900 Hz consistent with Doppler 
broadening \cite{dopplerfootnote}.
The response of trapped condensates was
strikingly different.
As the density of the trapped condensates was increased, 
the Bragg scattering resonance was significantly weaker in
strength and shifted upwards in frequency.
This reflects the changing character of the excitations created 
by Bragg scattering as the speed of sound was increased:
at a fixed Bragg scattering momentum, the excitations 
passed from the free-particle to the phonon regime.

To account for this behavior, we use
the  Bogoliubov description of the 
interacting homogeneous Bose-Einstein condensate \cite{bogo47}.
The hamiltonian
\begin{equation}
	{\mathcal H} = \sum_{k} \hbar \omega_k^0 \hat{a}^{\dagger}_k \hat{a}_k + 
\sum_{k, l, m} \frac{2 \pi \hbar^{2} a}{m V} \hat{a}^{\dagger}_k \hat{a}^{\dagger}_l 
\hat{a}_m \hat{a}_{k+l-m},
	\label{eq:hami}
\end{equation}
for a gas in volume V,
where $\hbar \omega^{0}_{k} = \hbar^2 k^2 / 2 m$,
is approximated by 
replacing the zero-momentum operators with $c$-numbers 
$\hat{a}^{\dagger}_0 = \hat{a}_0 = \sqrt{N_0}$
where $N_0$ is the number of atoms with zero momentum.
Neglecting terms of order $N^{-1/2}$, the hamiltonian is diagonalized by a canonical 
transformation to operators
defined by
$\hat{a}_k = u_k \hat{b}_k - v_k \hat{b}^{\dagger}_{-k}$
where $u_{k} = \cosh \phi_{k}$, 
$v_{k} = \sinh \phi_{k}$ and 
$\tanh 2 \phi_{k} = \mu / (\hbar \omega_{k}^{0} + \mu)$.
The energy of the Bogoliubov excitation
created by $\hat{b}^{\dagger}_{k}$ is 
$\hbar \omega_k^{B} = \sqrt{ \hbar \omega^{0}_{k} (\hbar \omega^{0}_{k} + 
2 \mu) }$.

Neglecting small contributions
representing multiparticle 
excitations \cite{nozi90,grif93},
the single quasi-particle contribution to the 
static structure factor is
\begin{equation}
S({\bf q}) = \frac{N_0}{N} \langle g | \! \left( \hat{a}_{q} 
\hat{a}^\dagger_{q} +
\hat{a}^\dagger_{-q} \hat{a} _{-q}  +
 \hat{a}^\dagger_{-q} \hat{a}^\dagger_{q} +
\hat{a} _{q} \hat{a} _{-q} \right) \! | g \rangle.
\end{equation}
Substituting the Bogoliubov operators, one obtains \cite{grahamref}
\begin{equation}
	S({\bf q}) \simeq (u_{q}^{2} + v_{q}^{2} - 2 u_{q} 
	v_{q})  = \omega_{q}^{0} / \omega_{q}^{B}.
	\label{eq:singleqs}
\end{equation}
In the limit $\hbar \omega_{q}^{0} \gg \mu$, the Bogoliubov excitations 
become identical to free-particle excitations ($u_{q} \rightarrow 1$,
$v_{q} \rightarrow 
0$), and $S({\bf q}) \rightarrow 1$. 
For phonons ($\hbar \omega_{q}^{0} \ll \mu$), $S({\bf q}) \rightarrow 
\hbar q/2 m c_{s} = q \xi / \sqrt{2}$ and the line strength diminishes
linearly with $q$.

To the same order of approximation, the quasi-particle resonance 
is undamped, and the dynamical structure factor is simply 
$S({\bf q}, \omega) = S({\bf q}) \delta( 
\omega - \omega_{q}^{B})$ (satisfying the $f$-sum rule: $\int \omega 
S({\bf q}, \omega) d \omega = \omega_{q}^{0}$ \cite{nozi90}).
Thus, accompanying the diminished line strength,
the Bragg resonance is shifted upward from the free particle 
resonance by $\omega_{q}^{B} - \omega_{q}^{0}$.

Equivalently, the suppression of the Bragg resonance in the phonon 
regime can be understood in terms of the many-body condensate 
wavefunction.
The static structure factor is the magnitude of the state vector 
$|e\rangle = \sum_{k} \hat{a}^{\dagger}_{k+q} \hat{a}_{k} |g\rangle / \sqrt{N}$.
The macroscopic population of the
zero-momentum state picks out two 
relevant terms in the summation:
\begin{equation}
	|e \rangle \simeq (\hat{a}^{\dagger}_{q} \hat{a}_{0} |g \rangle + 
	\hat{a}^{\dagger}_{0} \hat{a}_{-q} |g \rangle) / \sqrt{N} = 
	|e^{+}\rangle + |e^{-}\rangle
	\label{eq:evector}
\end{equation}
These represent two means by which momentum is 
imparted to the condensate: either by promoting a zero-momentum 
particle to momentum $\hbar {\bf q}$, or else by demoting a
particle from momentum $- \hbar {\bf q}$ to zero momentum.

If correlations could be neglected, the total rate of excitation 
would simply be the sum of the independent rates for these two 
processes, proportional to
$\langle e^{+}|e^{+} \rangle = \langle N_{q}\rangle + 1 = 
u_{q}^{2}$ and $\langle e^{-}|e^{-} \rangle = \langle N_{-q} 
\rangle = 
v_{q}^{2}$ where $\langle N_{k} \rangle$ is the expected number of 
atoms of momentum $\hbar {\bf k}$ in the condensate.
This would apply, for example, to a condensate in a pure number state,
or to an ideal gas condensate 
with a thermal admixture of atoms with momenta 
$\pm \hbar {\bf q}$, and would always lead to $S({\bf q}) > 1$.

Yet, for the many-body ground state of the interacting Bose gas, the 
behavior is dramatically different.
Collisions of zero-momentum atoms 
admix into the condensate pairs of atoms at momenta $\pm \hbar {\bf 
q}$, the population of which comprises the quantum depletion 
\cite{huan87}.
As a result, the two momentum transfer mechanisms 
described above produce indistinguishable states, and the 
rate of momentum transfer is given by the interference of two 
amplitudes, not by the sum of two rates.
Pair excitations in the condensate are correlated so as to minimize 
the total energy, and thereby give destructive interference between 
the two momentum transfer processes,
i.e.\ $S({\bf q}) = (u_{q} - v_{q})^{2} < 1$.
For high momentum,
$\langle N_{q}\rangle \ll 1$ and the interference plays a 
minor role.
In the phonon regime, while the independent rates $u_{q}^{2}$ and 
$v_{q}^{2}$ (and hence $\langle N_{\pm q} \rangle$)
diverge as $1/q$, the highly correlated quantum 
depletion 
extinguishes the rate of Bragg excitation.
% \cite{pairdiscussion}.

These results for the homogeneous Bose gas can be directly applied to
trapped, inhomogeneous condensates by 
a local density approximation
since the reduced phonon wavelength $q^{-1}$ (0.4 $\mu$m) is much 
smaller than the size of the condensate ($> 20 \, \mu$m) \cite{timmpc,csordasrefs}.
In the Thomas-Fermi regime, the condensate has a normalized density 
distribution $f(n) = 15 n / 4 n_{0} \sqrt{1 - n/n_{0}}$ where $n_{0}$ 
is the maximum condensate density.
The Bragg excitation line shape is then
\begin{equation}
	I(\omega) \, d\omega = \frac{15}{8} \frac{\omega^{2} - 
	{\omega_{q}^{0}}^{2}}{\omega_{q}^{0} (\mu / \hbar)^{2}} 
	\sqrt{1 - \frac{\omega^{2} - 
	{\omega_{q}^{0}}^{2}}{2 \omega_{q}^{0} \mu / \hbar}} \, d\omega
	\label{eq:lineshape}
\end{equation}
which can be integrated to obtain the line strength $S({\bf q})$ and center 
frequency.
The line strength has the limiting values of 
$S({\bf q}) \rightarrow 15 \pi/32 \, (\hbar \omega_{q}^{0}/ 
2 \mu)^{1/2}$ in the phonon regime and $S({\bf q}) \rightarrow 1 - 4 \mu / 7 
\hbar \omega_{q}^{0}$ in the free-particle regime \cite{explicitform}.
In accordance with the $f$-sum rule, the center frequency 
$\bar{\omega}$ is given as $\omega_{q}^{0} / S({\bf q})$.

These predictions are shown in Fig.\ 
\ref{sqshift} using $\omega_{0}^{q} = 2 \pi \times$ 
1.38 kHz which accounts for the expected mean-field shift of the 
Bragg resonance for the expanded condensate.
Both the line strength and the shift of the Bragg resonance are well 
described by our treatment.
For comparison, previous measurements \cite{sten99brag}
of the mean-field shift of the Bragg 
resonance ($4 \mu / 7 \hbar$) in the 
free-particle regime are also shown, clearly indicating the many-body 
character of low energy excitations.

In conclusion, stimulated light scattering was used to excite phonons 
in trapped Bose-Einstein condensates with wavelengths much smaller 
than the size of the trapped sample.
The static structure factor was shown to be 
substantially reduced in the phonon regime.
This modification of light-atom interactions arises from the 
presence of a correlated admixture of momentum excitations in the 
condensate.
The observed reduction of $S({\bf q})$ also implies a reduction
of inelastic Rayleigh scattering of light with 
wavevector $k$ by a condensate when $\hbar \omega_{k}^{0} < \mu$ \cite{suppression}.
This effect may reduce heating in optical dipole traps and reduce the 
optical density probed in absorption imaging.
For example,
the absorption of near-resonant light 
by a homogeneous sodium condensate 
at a density of $3 \times 10^{15} \, \mbox{cm}^{-3}$ \cite{stam98odt}
should be reduced by a factor of 
two.

This work was supported by the Office of Naval Research, 
NSF, JSEP, ARO, NASA, and the David and Lucile
Packard
Foundation.
A.P.C.\  acknowledges additional support from the NSF, 
A.G.\ from DAAD, and D.M.S.-K.\
from JSEP.

\bibliographystyle{prsty}
\bibliography{wkrefs,structurerefs}

\begin{thebibliography}{10}

\bibitem{grey78}
T.J. Greytak,  in {\em Quantum Liquids}, edited by J. Ruvalds and T. Regge
  (North-Holland, New York, 1978).

\bibitem{soko95}
P.E. Sokol,  in {\em Bose-Einstein Condensation}, edited by A. Griffin, D.W.
  Snoke, and S. Stringari (Cambridge University Press, Cambridge, 1995), p.\
  51.

\bibitem{nozi90}
P. Nozi{\`e}res and D. Pines, {\em The Theory of Quantum Liquids}
  (Addison-Wesley, Redwood City, CA, 1990).

\bibitem{grif93}
A. Griffin, {\em Excitations in a Bose-condensed liquid} (Cambridge University
  Press, Cambridge, 1993).

\bibitem{sten99brag}
J. Stenger {\it et~al.}, Phys. Rev. Lett. {\bf 82},  4569  (1999).

\bibitem{earlyrefs}
D.S.\ Jin {\it et al.}, Phys.\ Rev.\ Lett.\ {\bf 77}, 420 (1996); M.-O.\ Mewes
  {\it et al.}, Phys.\ Rev.\ Lett.\ {\bf 77}, 988 (1996); D.S.\ Jin {\it et
  al.}, Phys.\ Rev.\ Lett.\ {\bf 78}, 764 (1997); M.R. Andrews {\it et~al.},
  Phys. Rev. Lett. {\bf 79}, 553 (1997); D.M.\ Stamper-Kurn {\it et al.},
  Phys.\ Rev.\ Lett.\ {\bf 81}, 500 (1998).

\bibitem{cohe92}
C. Cohen-Tannoudji, J. Dupont-Roc, and G. Grynberg, {\em Atom-Photon
  Interactions} (Wiley, New York, 1992).

\bibitem{java95scat}
J. Javanainen and J. Ruostekoski, Phys. Rev. A {\bf 52},  3033  (1995).

\bibitem{mewe96bec}
M.-O. Mewes {\it et~al.}, Phys. Rev. Lett. {\bf 77},  416  (1996).

\bibitem{inou99super}
S. Inouye {\it et~al.}, 1999, submitted.

\bibitem{cast96}
Y. Castin and R. Dum, Phys. Rev. Lett. {\bf 77},  5315  (1996).

\bibitem{fallingfootnote}
The laser beams had diameters of about 2 mm and were directed vertically, so
  the 120 $\mu$m drop of the atomic sample during the 5 ms time of flight was
  negligible.

\bibitem{kett99vartemp}
W. Ketterle, D.S. Durfee, and D.M. Stamper-Kurn, in {\em Bose-Einstein
  Condensation in Atomic Gases}, {\em Proceedings of the International School
  of Physics ``Enrico Fermi''}, edited by M. Inguscio, S. Stringari, and C.E.
  Wieman, to be published.

\bibitem{sqwfootnote}
The momentum transfer is proportional to $S({\bf q}, \omega) - S(-{\bf q},
  \omega)$.

\bibitem{residualfootnote}
The Bragg resonance frequency of the expanding condensate includes an expected
  160 Hz mean-field shift.

\bibitem{dopplerfootnote}
A tilt angle of $\approx 5^{\circ}$ of ${\bf q}$ with respect to the axial
  direction introduced Doppler broadening from the radial expansion.

\bibitem{bogo47}
N.N. Bogoliubov, J. Phys. (USSR) {\bf 11},  23  (1947).

\bibitem{grahamref}
Shown explicitly for light scattering from a Bose-Einstein condensate in R.\
  Graham and D.\ Walls, Phys.\ Rev.\ Lett.\ {\bf 76}, 1774 (1996).

\bibitem{huan87}
K. Huang, {\em Statistical Mechanics} (Wiley, New York, 1987).

\bibitem{timmpc}
E.\ Timmermans and P.\ Tommasini, private communication.

\bibitem{csordasrefs}
The discrete spectrum was considered by A.\ Csord\'{a}s, R.\ Graham, and P.\
  Sz\'{e}pfalusy, Phys.\ Rev.\ A {\bf 54}, R2543 (1996); {\it ibid.}\ {\bf 57},
  4669 (1998).

\bibitem{explicitform}
$S({\bf q}) = 15 \eta /64 (y + 4 \eta - 2 y \eta^2 + 12 \eta^3 - 3 y \eta^4)$
  where $\eta^2 = \hbar \omega_q^0 / 2 \mu$ and $y = \pi - 2 \arctan ((\eta^2 -
  1)/2 \eta)$.

\bibitem{suppression}
For a homogeneous Bose-Einstein condensate, inelastic scattering is reduced by
  a factor $(\sqrt{x (1 + x)} - \sinh^{-1} \! \sqrt{x}) / x$ where $x = 2 \hbar
  \omega_{k}^{0} / \mu$.

\bibitem{stam98odt}
D.M. Stamper-Kurn {\it et~al.}, Phys. Rev. Lett. {\bf 80},  2072  (1998).

\end{thebibliography}

\end{document}